\documentclass{aa} %   standard A&A printer layout

\usepackage{graphicx}
\usepackage[varg]{txfonts}  %required by A&A

\usepackage{amsmath, amssymb, amsfonts}

\usepackage{color}

\usepackage[pagewise]{lineno}

\begin{document}
% \linenumbers	%number the lines pagewise
% \pagewiselinenumbers	%number the lines consecutively

\title{Spatial damping of propagating sausage waves in coronal cylinders}
\author{Ming-Zhe Guo\inst{\ref{inst1}, \ref{inst2}}
	\and
	Shao-Xia Chen\inst{\ref{inst1}}
	\and
	Bo Li\inst{\ref{inst1}}
	\and
	Li-Dong Xia\inst{\ref{inst1}}
	\and
	Hui Yu\inst{\ref{inst1}}
	}

%\offprints{B. Li, \email{bbl@sdu.edu.cn}}

\institute{Shandong Provincial Key Laboratory of Optical Astronomy and
	  Solar-Terrestrial Environment, and Institute of Space Sciences,
	  Shandong University Weihai, Weihai 264209, China \\
      \email{bbl@sdu.edu.cn}\label{inst1}
      \and
      CAS Key Laboratory of Geospace Environment, University of Science \& Technology of China, Chinese Academy of Sciences, Hefei 230026, China
	\label{inst2}
      }

%\date{Received date / Accepted date }
\date{} % leave blank for now
\abstract
{
Sausage modes are important in coronal seismology. 
   Spatially damped propagating sausage waves were recently observed in the solar atmosphere.
}
{
We examine how wave leakage influences the spatial damping of sausage waves propagating along
   coronal structures modeled by a cylindrical density enhancement
   embedded in a uniform magnetic field.
}
{
Working in the framework of cold magnetohydrodynamics,
   we solve the dispersion relation (DR) governing sausage waves for complex-valued longitudinal wavenumber $k$
   at given real angular frequencies $\omega$.
For validation purposes, we also provide analytical approximations to the DR in the low-frequency limit 
    and in the vicinity of $\omega_{\rm c}$, the critical angular frequency separating trapped from leaky waves.
    }
{
In contrast to the standing case, propagating sausage waves are allowed for $\omega$ much lower than $\omega_{\rm c}$.
However, while able to direct their energy upwards, these low-frequency waves are subject to substantial spatial attenuation.
The spatial damping length shows little dependence on the density contrast between the cylinder and its surroundings,
   and depends only weakly on frequency.
This spatial damping length is of the order of the cylinder radius for $\omega \lesssim 1.5 v_{\rm Ai}/a$, 
   where $a$ and $v_{\rm Ai}$ are the cylinder radius and the Alfv\'en speed
   in the cylinder, respectively.
   }
{
If a coronal cylinder is perturbed by symmetric boundary drivers (e.g., granular motions) with a broadband spectrum,
    wave leakage efficiently filters out the low-frequency components.
    }
\maketitle

\titlerunning{Propagating sausage waves in coronal cylinders}
\authorrunning{Guo et al.}

\section{Introduction}
\label{sec_intro}
Considerable progress has been made in coronal seismology thanks to the abundantly identified waves
    and oscillations in the structured solar atmosphere
    (for a recent review,
    see~\citeauthor{2012RSPTA.370.3193D}~\citeyear{2012RSPTA.370.3193D},
    and also~\citeauthor{2007SoPh..246....1B}~\citeyear{2007SoPh..246....1B},
    \citeauthor{2009SSRv..149....1N}~\citeyear{2009SSRv..149....1N},
    \citeauthor{2011SSRv..158..167E}~\citeyear{2011SSRv..158..167E} for three recent topical issues).
Equally important is a detailed theoretical understanding of the collective
    wave modes supported by magnetized cylinders~\citep[e.g.,][]{2000SoPh..193..139R}.
While kink waves (with azimuthal wavenumber $m=1$) have attracted much attention
    since their measurements with TRACE~\citep{1999Sci...285..862N,1999ApJ...520..880A}, 
    sausage waves prove important in interpreting second-scale quasi-periodic pulsations (QPPs)
    in the lightcurves of solar flares~\citep[see][for a recent review]{2009SSRv..149..119N}.
Their importance is strengthened given their {\bf recent detection} in both the chromosphere~\citep{2012NatCo...3E1315M}
    and the photosphere~\citep{2014A&A...563A..12D,2015ApJ...806..132G}.

Standing sausage modes are well understood.
For instance, two distinct regimes are known to exist, depending on
    the longitudinal wavenumber $k$~\citep{2005LRSP....2....3N}.
The trapped regime results when $k$ exceeds some critical value $k_{\rm c}$, 
    whereas the leaky regime 
    arises when the opposite is true.
Both eigen-mode analyses~\citep{2007AstL...33..706K,2014ApJ...781...92V} 
    and numerical simulations from an initial-value-problem perspective~\citep[e.g.,][]{2012ApJ...761..134N}
    indicate that the period $P$ of sausage modes increases smoothly with decreasing $k$ 
    until reaching some $P_0$ in the thin-cylinder limit
    ($ka\rightarrow 0$ with $a$ being the cylinder radius).
Likewise, being identically infinite in the trapped regime, the attenuation time $\tau$ decreases
    with decreasing $k$ before saturating at $\tau_0$ when $ka =0$.
Furthermore, $P_0$ is determined primarily by $a/v_{\rm Ai}$, where $v_{\rm Ai}$ is the 
    Alfv\'en speed in the cylinder~\citep{1984ApJ...279..857R}, with the detailed transverse density distribution 
    playing {\bf an important role (\citeauthor{2012ApJ...761..134N}~\citeyear{2012ApJ...761..134N},
    also \citeauthor{2015arXiv150702169C}~\citeyear{2015arXiv150702169C})}. 
This is why second-scale QPPs are attributed to standing sausage modes, since $a/v_{\rm Ai}$ is of the order
    of seconds for typical coronal structures.
On the other hand, the ratio $\tau_0/P_0$ is basically proportional to the density contrast~\citep[e.g.,][]{2007AstL...33..706K},
    meaning that high-quality sausage modes are associated with coronal structures with densities
    considerably exceeding their surroundings.

Interestingly, the dispersive behavior of trapped modes is important also in understanding impulsively
    generated sausage waves~\citep{1984ApJ...279..857R}.
When measured at a distance from the source, the signals from these propagating waves possess three phases:
    periodic, quasi-periodic, and decay.
The frequency dependence of the longitudinal group speed $v_{\rm gr}$ is crucial in this context.
In particular, whether the quasi-periodic phase exists depends on the existence of a local minimum in $v_{\rm gr}$,
    which in turn depends on the density profile transverse
    to the structure~\citep{1988A&A...192..343E, 1995SoPh..159..399N}.
This analytical expectation, extensively reproduced in numerical
    simulations~\citep[e.g.,][]{1993SoPh..144..101M,2004A&A...422.1067S,2004MNRAS.349..705N},
    well explains the time signatures of the wave trains discovered with the Solar Eclipse Coronal
    Imaging System~\citep{2001MNRAS.326..428W,2002MNRAS.336..747W,2003A&A...406..709K}
    {\bf as well as those measured with SDO/AIA~\citep{2013A&A...554A.144Y}}.

We intend to examine the spatial damping of leaky sausage waves propagating
    along coronal cylinders in response to photospheric motions due to, say, granular convection~\citep{1996ApJ...472..398B}.
One reason for conducting this study is that, 
    besides the observations showing that propagating sausage waves abound
    in the chromosphere~\citep{2012NatCo...3E1315M},
    a recent study clearly demonstrates the spatial damping of sausage waves propagating from
    the photosphere to the transition region in a pore~\citep{2015ApJ...806..132G}.
Another motivation is connected to the intensive interest~\citep{2010A&A...524A..23T,2013A&A...551A..39H,2013A&A...551A..40P}
    in employing resonant absorption to understand the spatial damping of propagating
    kink waves measured with the Coronal Multi-Channel Polarimeter (CoMP) instrument~\citep{2007Sci...317.1192T,2009ApJ...697.1384T}.
A leading mechanism for interpreting the temporal damping of standing
    kink modes~\citep[][and references therein]{2002ApJ...577..475R, 2003ApJ...598.1375A},
    resonant absorption is found to attenuate propagating kink waves with
    a spatial length inversely proportional to wave frequency~\citep{2010A&A...524A..23T}.
If attributing the generation of these kink waves to broadband photospheric perturbations,
    one expects that resonant absorption essentially filters out the high-frequency components.
One then naturally asks: What role does wave leakage play in attenuating propagating sausage waves?
    And what will be the frequency dependence of the associated damping length?

This manuscript is structured as follows.
We present the necessary equations, the dispersion relation (DR) in particular, in Sect.\ref{sec_model}, 
   and then present our numerical solutions to the DR in Sect.\ref{sec_numres}
   where we also derive a couple of analytical approximations to the DR for validation purposes.
Finally, a summary is given in Sect.\ref{sec_conc}.

\section{Problem Formulation}
\label{sec_model}

We consider sausage waves propagating in a structured corona modeled by a plasma cylinder with radius $a$
   embedded in a uniform magnetic field ${\bf B}=B\hat{z}$,
   where a cylindrical coordinate system $(r,\theta,z)$ is adopted.  
The cylinder is directed along the $z$-direction.
A piece-wise constant (top-hat) density profile is adopted, with the densities inside and external to
   the cylinder being $\rho_{\rm i}$ and $\rho_{\rm e}$, respectively ($\rho_{\rm e} < \rho_{\rm i}$).
The Alfv\'en speeds, $v_{\rm Ai}$ and $v_{\rm Ae}$, follow from the definition $v_{\rm A} = \sqrt{B^2/4\pi\rho}$.
Appropriate for the solar corona, zero-$\beta$, ideal MHD equations are adopted.  
In such a case, sausage waves do not perturb the $z$-component of the plasma velocity.
Let $\delta v_r$ denote the radial velocity perturbation, and 
      $\delta b_r$, $\delta b_z$ denote the radial and longitudinal components of the perturbed magnetic field $\delta \vec{b}$,
      respectively.
The perturbed magnetic pressure, or equivalently total pressure in the zero-$\beta$ case,
     is then $\delta p_{\mathrm{tot}} = B\delta b_z/4\pi$.

Let us Fourier-decompose any perturbed value $\delta f(r, z;t)$ as 
\begin{eqnarray}
\label{eq_Fourier}
  \delta f(r,z;t)={\rm Re}\left\{\tilde{f}(r)\exp\left[-i\left(\omega t-kz\right)\right]\right\}~.
\end{eqnarray}
{\bf With the definition
\begin{eqnarray}
\label{eq_mu}
  \mu_{\rm i}^2=\frac{\omega^2}{v_{\rm Ai}^2}-k^2, \hspace{0.2cm}
  \mu_{\rm e}^2=\frac{\omega^2}{v_{\rm Ae}^2}-k^2  \hspace{0.2cm}
  (-\frac{\pi}{2} < \arg{\mu}_{\rm i}, \arg{\mu}_{\rm e} \le \frac{\pi}{2}) ,
\end{eqnarray}
   linearizing the ideal MHD equations then leads to~\citep[e.g.,][]{1986SoPh..103..277C}
}
\begin{eqnarray}
\label{eq_ptilde}
   \frac{1}{r}\left(r\tilde{p}'_{\rm tot}\right)'
    + \mu^2 \tilde{p}_{\rm tot} = 0 , 
\end{eqnarray}
   where the prime $' = d/dr${\bf, and this equation is valid
   both inside (denoted by the subscript i) and outside (the subscript e) the cylinder.}
The solutions to Eq.~(\ref{eq_ptilde}) are given by 
\begin{eqnarray}
\tilde{p}_{\rm tot,i} = A_{\rm i} J_0(\mu_{\rm i} r),
\tilde{p}_{\rm tot,e} = A_{\rm e} H_0^{(1)}(\mu_{\rm e} r),
\end{eqnarray}
   where  $J_n$ and $H_n^{(1)}$ are the $n$-th-oder Bessel and Hankel functions of
   the first kind, respectively (here $n=0$).
For future reference, we also give the expressions for the Fourier amplitudes 
   of some relevant perturbations,
\begin{eqnarray}
\label{eq_Fourier_amp}
   \tilde{b}_z = \frac{4\pi \tilde{p}_{\rm tot}}{B}, 		\hskip 0.1cm
   \tilde{v}_r = \frac{-i \omega}{\mu^2}\frac{\tilde{b}_z'}{B},	\hskip 0.1cm
   \mbox{and}	\hskip 0.1cm
   \tilde{b}_r = -\frac{k B}{\omega} \tilde{v}_r .
\end{eqnarray}
In addition, the energy flux density carried by the sausage waves is given by
\begin{eqnarray}
\label{eq_eflx}
\vec{F} = -\frac{1}{4\pi}\left(\delta \vec{v}\times \vec{B}\right)\times \delta \vec{b}
   = \delta p_{\mathrm{tot}}\delta v_r \hat{r} - \frac{\delta v_r \delta b_r}{4\pi}B\hat{z} . 
\end{eqnarray}

The dispersion relation (DR) for sausage waves follows from the requirements that
    {\bf the Fourier amplitudes of both the total pressure perturbation $\tilde{p}_{\rm tot}$
    and the radial velocity perturbation $\tilde{v}_r$}
    be continuous at the
    cylinder boundary $r=a$.
{\bf Note that more properly, the continuity of the Fourier amplitude of the radial Lagrangian displacement $\tilde{\xi}_r$
    should be used in place of that of $\tilde{v}_r$.
Even though in the static case the two requirements are equivalent given that $\tilde{v}_r = -i\omega\tilde{\xi}_r$,
    the equivalence will not be present when axial flows exist
    \citep[e.g.,][]{1992SoPh..138..233G, 2014A&A...568A..31L}.}
The DR reads~\citep[e.g.,][]{1986SoPh..103..277C}
\begin{eqnarray}
\label{eq_DR}
\frac{\mu_{\rm i}}{\mu_{\rm e}}=\frac{J_1(\mu_{\rm i}a)}{J_0(\mu_{\rm i}a)}
\frac{H^{(1)}_0(\mu_{\rm e}a)}{H^{(1)}_1(\mu_{\rm e}a)}.
\end{eqnarray}
Throughout this study, we examine only the solutions that correspond to the lowest
    critical longitudinal wavenumber in the trapped regime.
Unless otherwise specified, we solve Eq.~(\ref{eq_DR}) by assuming that
    the angular frequency $\omega$ is real,
    and the longitudinal wavenumber $k$ is complex ($k = k_{\rm R} + i k_{\rm I}$).
For these propagating waves, the energy flux density averaged over a wave period $2\pi/\omega$ is
\begin{eqnarray}
&& \left< F_r \right> = \frac{1}{2}\mathrm{Re}\left(\tilde{p}_{\rm tot} \tilde{v}_r^*\right) \exp(-2 k_{\rm I} z) , 
  \label{eq_eflx_r} \\
&& \left< F_z \right> = \frac{1}{2}\mathrm{Re}\left( -\frac{B}{4\pi}\tilde{b}_r\tilde{v}_r^*\right) \exp(-2 k_{\rm I} z) ,
  \label{eq_eflx_z}
\end{eqnarray}
    where the subscripts $r$ and $z$ denote the radial and longitudinal components, respectively.
Furthermore, $\tilde{f}^*$ means taking the complex conjugate of some complex-valued $\tilde{f}$.
{\bf (See Appendix~A for a derivation of Eqs.~(\ref{eq_eflx_r}) and (\ref{eq_eflx_z}).)}

{\bf Before proceeding, we note that a comprehensive study was conducted recently
   by~\citet[][hereafter M15]{2015A&A...578A..60M}
   to work out the specific expressions for the energy and energy flux densities
   for both fast and slow sausage waves. 
That study differs from ours primarily in the objectives.
To enable a calculation of the energy content of
   sausage waves in a variety of solar structures, M15 employed
   the specific expressions for the eigen-functions and focused on trapped waves.
However, the aim of the present study is to examine the spatial damping of propagating leaky waves.
For this purpose, the specific expressions for the energy flux densities are not necessary.
Rather, we only need to evaluate the sign of $\left< F_r \right>$ at large distances from
   the cylinder to offer a physical explanation for the spatial damping.
Likewise, the sign of $\left< F_z \right>$ is necessary to address whether the spatially damped waves
   can impart their energy upwards.
We would like to stress that, while the energetics of leaky waves cannot be addressed with an eigen-value-problem approach
   (see our discussion towards the end of Sect.~\ref{sec_numres}),
   evaluating the signs of $\left< F_r \right>$ and $\left< F_z \right>$ makes physical sense.
A similar conclusion was reached in M15 for the trapped waves at the critical axial wavenumber,
   below which trapped waves are no longer allowed.
Regarding the technical details, in the present study the energy flux density, Eq.~(\ref{eq_eflx}), is equivalent to
   the Poynting flux given the cold MHD approximation.
In other words, the contribution due to thermal pressure, given by $\left<\vec{T}\right>$ in Eq.~(4) of M15,
   vanishes here.
}

\section{Results}
\label{sec_numres}
\begin{figure}
\centering
\includegraphics[width=0.95\hsize]{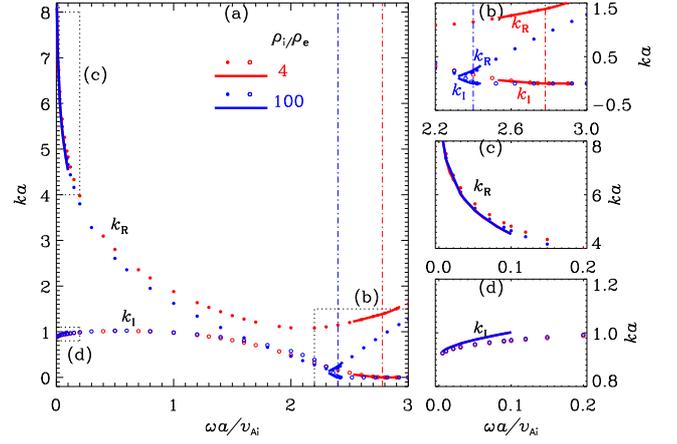}
 \caption{Dependence of the real ($k_{\rm R}$, the solid dots) and imaginary ($k_{\rm I}$, open dots)
     parts of the longitudinal wavenumber on the angular frequency $\omega$.
 Two density ratios $\rho_{\rm i}/\rho_{\rm e} = 4$ and $100$ are examined, and are given
     by the symbols and curves in red and blue, respectively.
 The vertical dash-dotted lines correspond to the critical angular frequency $\omega_{\rm c}$ that
     separates the leaky (to the left of $\omega_{\rm c}$) from trapped (right) regimes.
 Figure~\ref{fig_vali}a presents an overview of the dispersion curves,
     while Fig.\ref{fig_vali}b (Figs.~\ref{fig_vali}c and \ref{fig_vali}d)
     examines the portion where $\omega$ is close to $\omega_{\rm c}$ ($\omega$ approaches zero).
 The dots are found by numerically solving the full dispersion relation (DR, Eq.~(\ref{eq_DR}))
     by assuming a real $\omega$ but a complex-valued $k=k_{\rm R} + i k_{\rm I}$,
     while the curves represent the solutions to the approximate DR (see text for details).
 }
\label{fig_vali}
 \end{figure}

Figure~\ref{fig_vali} presents the solutions to the DR for two different density ratios, 
    one mild ($\rho_{\rm i}/\rho_{\rm e} =4$, the red curves
    and symbols)
    and the other rather large ($\rho_{\rm i}/\rho_{\rm e} = 100$, blue).
The real (the filled dots) and imaginary (open) parts of the longitudinal wavenumber $k$ are presented
    as functions of the angular frequency $\omega$.
The vertical dash-dotted lines separate the trapped regime (where $k_{\rm I} \equiv 0$)
    from the leaky one ($k_{\rm I} \ne 0$), and correspond to
    $\omega_{\rm c} = k_{\rm c} v_{\rm Ae}$ 
    with $k_{\rm c} = 2.4048/(a\sqrt{\rho_{\rm i}/\rho_{\rm e}-1})$~\citep{2005LRSP....2....3N}.
Let us leave the physical interpretation of the results till later,
    and for now provide a validation study on the numerical solutions.
To this end, the dispersion curves in the neighborhood of $\omega_{\rm c}$
    are amplified in Fig.~\ref{fig_vali}b.
Assuming that $|\Delta \omega| = |\omega - \omega_{\rm c}| \ll \omega_{\rm c}$, 
    one can derive an analytical expression, 
\begin{eqnarray}
 \label{eq_kc}
 \Delta k = 
    \frac{1+\frac{i}{\pi}
     \left\{\ln\left[\frac{k^2_{\rm c}a^2}{2}\left(\frac{\Delta \omega}{\omega_{\rm c}}-\frac{\Delta k}{k_{\rm c}}\right)\right]
         +2\gamma-\frac{v_{\rm Ae}^2}{v_{\rm Ai}^2}\right\}}
    {1+\frac{i}{\pi}
     \left\{\ln\left[\frac{k^2_{\rm c}a^2}{2}\left(\frac{\Delta \omega}{\omega_{\rm c}}-\frac{\Delta k}{k_{\rm c}}\right)\right]
         +2\gamma-1\right\}}
    \frac{k_{\rm c}}{\omega_{\rm c}}\Delta \omega ,
 \end{eqnarray}
   where $\Delta k = k-k_{\rm c}$.
Interestingly, Eq.~(\ref{eq_kc}) agrees with Eq.~(16) in~\citet{2014ApJ...781...92V} even though standing modes
    with real $k$ and complex-valued $\omega$ are examined there.
The primary difference between the two is
    the appearance of the Euler constant $\gamma = 0.577$, which we found is necessary to retain
    when the logarithmic term does not substantially exceed unity.
In Fig.~\ref{fig_vali}b, the solid curves represent the solutions to Eq.~(\ref{eq_kc}).
It can be seen that they well approximate the solutions to the full DR given by the dots
    in both trapped and leaky regimes.

Another portion of the dispersion curves that is analytically tractable is when $\omega \rightarrow 0$.
In this case, the DR can be approximated by 
\begin{eqnarray}
\label{eq_smallomega}
\mu_{\rm i}a = {\rm arctan}\left(\frac{\mu_{\rm e}-i\mu_{\rm i}}{\mu_{\rm e} +i\mu_{\rm i}}\right)~.
\end{eqnarray}
To arrive at Eq.~(\ref{eq_smallomega}), we have assumed that $|\mu_{\rm i} a|, |\mu_{\rm e} a| \gg 1$,
    which can be justified a posteri.
The real (imaginary) part of the solution to Eq.~(\ref{eq_smallomega}) 
    is presented by the curves in Fig.~\ref{fig_vali}c (Fig.~\ref{fig_vali}d),
    and is found to agree well with the solutions to the full DR, represented by the dots.
It is also clear that  
    in the low-frequency portion the solutions to the DR for the two density ratios are close to each other,
    despite that the ratios differ substantially.
This is understandable from Eqs.~(\ref{eq_smallomega}) and (\ref{eq_mu}), 
    because $\mu_{\rm i}^2$ and $\mu_{\rm e}^2$ both approach $-k^2$ when $\omega$ approaches zero,
    therefore having little dependence on the Alfv\'en speeds or densities.

\begin{figure}
\centering
\includegraphics[width=0.95\hsize]{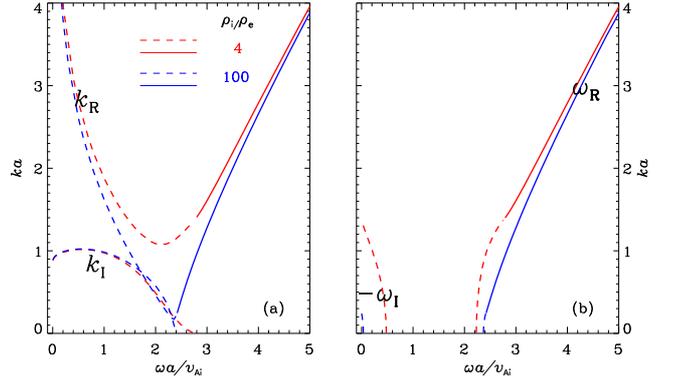}
 \caption{Dispersion curves of (a) propagating and (b) standing sausage waves.
 In (a) the real ($k_{\rm R}$) and imaginary ($k_{\rm I}$)
    parts of the longitudinal wavenumber are given as functions of the real
    angular frequency $\omega$.
 In (b) the real ($\omega_{\rm R}$) and imaginary ($\omega_{\rm I}$) parts of
    $\omega$ are shown as functions of the real wavenumber $k$, which is
    given by the vertical axis though.
 In both panels, the solid (dashed) curves represent the trapped (leaky) regime.
 Two density ratios ($\rho_{\rm i}/\rho_{\rm e} = 4$ and $100$) are examined
    as represented by the red and blue curves, respectively.
}
\label{fig_prop_vs_stand}
\end{figure}

Figure~\ref{fig_prop_vs_stand}a presents once again the dispersion curves for propagating sausage waves,
    where the real and imaginary parts of $k$ are given as functions of $\omega$.
For comparison, Fig.~\ref{fig_prop_vs_stand}b presents the dispersion curves for standing waves, 
    obtained by solving Eq.~(\ref{eq_DR}) for complex-valued $\omega$ ($\omega = \omega_{\rm R} + i \omega_{\rm I}$) 
    at given real values of $k$.
Note that here $k$ is given by the vertical axis, and $-\omega_{\rm I}$ is plotted instead of $\omega_{\rm I}$
    because $\omega_{\rm I} \le 0$.    
In both cases, we examine two density ratios with $\rho_{\rm i}/\rho_{\rm e}$ being $4$ (the red curves)
    and $100$ (blue), and present the trapped (leaky) regime with the solid (dashed) curves.
One can see that in the trapped regime, Figs.~\ref{fig_prop_vs_stand}a and \ref{fig_prop_vs_stand}b 
    agree exactly with one another, as expected for the situation where both $\omega$ and $k$ are real.
In the trapped regime, however, some distinct differences appear.
For standing waves, with decreasing $k$, the real (imaginary) part of the angular frequency $\omega_{\rm R}$ ($\omega_{\rm I}$) 
   decreases (increases in magnitude) and saturates when $k$ approaches zero.
This $k$-dependence of both $\omega_{\rm R}$ and $\omega_{\rm I}$ is well
    understood~\citep[e.g.,][]{2007AstL...33..706K,2012ApJ...761..134N,2014ApJ...781...92V}.
It suffices to note that
    the temporal damping clearly depends on $\rho_{\rm i}/\rho_{\rm e}$: the higher
    the density ratio, the slower the temporal damping.
In fact, for large $\rho_{\rm i}/\rho_{\rm e}$, a simple but reasonably accurate expression
    exists, namely, $|\omega_{\rm I}|/\omega_{\rm R} \approx (\pi/2)(\rho_{\rm e}/\rho_{\rm i})$~\citep{2007AstL...33..706K}.
When examining propagating waves, Fig.~\ref{fig_prop_vs_stand}a shows that the DR permits
     waves with $\omega$ much lower than $\omega_{\rm c}$, in contrast to the standing case.
In addition, for $\omega \lesssim v_{\rm Ai}/a$, neither $k_{\rm R}$ nor $k_{\rm I}$
    shows a significant dependence on the density ratio.
This is particularly true for $k_{\rm I}$, and has been explained in view of Eq.~(\ref{eq_smallomega}).

{\bf Why should the results be different if one simply changes from one perspective, where a DR
   is solved for complex $\omega$ as a function of real $k$, to another standpoint where the same DR
   is solved for complex $k$ as a function of real $\omega$?   
As has been discussed in detail by~\citet[][hereafter TFS95]{1995A&A...299..940T}, 
   indeed the two perspectives have a close relationship when wave attenuation is weak or absent.
However, some considerable difference between the two may result when strong attenuation takes place. 
An example for this can be found in~TFS95 where the authors examined linear Alfv\'en waves in a partially ionized gas
   where ions and neutrals are imperfectly coupled.
Solving the relevant DR for complex $\omega$ at real $k$ yields that $\omega$ may be purely imaginary in certain ranges of $k$,
   in other words, the waves may be overdamped and non-propagating.
In contrast, solving the same DR for complex $k$ at real $\omega$ yields that the real part of $k$ is always non-zero, meaning that
   the waves are always propagating.
This led TFS95 to conclude that how to choose a perspective depends on how the waves are excited.
One chooses complex $k$ and real $\omega$ to examine the spatial variation of the waves
   excited at a given location with a given real frequency.
On the other hand, one chooses complex $\omega$ and real $k$ to follow the temporal variation of the waves
   in response to perturbations initially imposed in a coherent way over many wavelengths.
These discussions are also valid if one examines the differences in Figs.~\ref{fig_prop_vs_stand}a and \ref{fig_prop_vs_stand}b.  
In particular, the absence of standing waves with frequencies below a certain value (see Fig.~\ref{fig_prop_vs_stand}b)
   can be understood given that in view of Fig.~\ref{fig_prop_vs_stand}a, 
   an initial perturbation cannot stay coherent in a longitudinal spatial range spanning many wavelengths
   and hence many damping lengths.
Instead, the system will select the frequencies and damping rates corresponding to the spatial periods 
   enforced externally.
}

Let us now focus on Fig.~\ref{fig_prop_vs_stand}a, where one can see that for both density ratios,
   once in the leaky regime, $k_{\rm I}$ increases with decreasing $\omega$ and
   somehow levels off when $\omega \lesssim 1.5 v_{\rm Ai}/a$.
On the other hand, $k_{\rm R}$ decreases when $\omega$ decreases from $\omega_{\rm c}$
   before increasing monotonically when $\omega$ further decreases. 
% Incidentally, the value for $\omega$ where $k_{\rm R}$ reaches a local minimum ($d k_{\rm R}/d \omega = 0$)
%     almost coincides with $\omega_{\rm R} (k=0)$ in Fig.~\ref{fig_prop_vs_stand}b where $d\omega_{\rm R}/d k = 0$. 
Let $\omega_{\rm m}$ denote the point where $k_{\rm R}$ reaches a local minimum.
It then follows that the apparent group speed $d \omega/d k_{\rm R} \le 0$  for $\omega \le \omega_{\rm m}$.
%     similar to what was found in \citet{1985ApJ...291..328D} where
%     coronal holes were examined as a leaky waveguide for fast waves.
One may question whether the sausage waves in this low-frequency portion can impart energy upwards.
As discussed in \citeauthor{1960Brillouin} (\citeyear{1960Brillouin}, chapter V), 
    when waves are heavily damped, the apparent group velocity may not represent 
    the velocity at which energy propagates.
In this case, we may directly use Eq.~(\ref{eq_eflx_z}) to evaluate the $z$-component of
    the wave energy flux density.
In view of Eq.~(\ref{eq_Fourier_amp}), one finds that
\begin{eqnarray*}
  \left< F_z \right> = \frac{B^2}{8\pi}\left|\tilde{v}_r\right|^2\left(\frac{k_{\rm R}}{\omega}\right) \exp(-2 k_{\rm I} z) ,
\end{eqnarray*}
   which is positive for positive $k_{\rm R}/\omega$, meaning that
   the low-frequency waves in question can still direct their energy upwards. 
However, in this case the plasma cylinder is such an inefficient waveguide
   that the wave energy is attenuated over  
   a longitudinal distance of the order of the cylinder radius.
   
More insights can be gained by further comparing the dispersive behavior of
   leaky standing and propagating sausage waves as given in Fig.~\ref{fig_prop_vs_stand}.
To start, let us note that for standing waves with real $k$ and complex $\omega$
   (propagating waves with real $\omega$ and complex $k$)
   the radial energy flux density in the external medium, 
   when averaged over a longitudinal wavelength $2\pi/k$ (a wave period $2\pi/\omega$), evaluates to
\begin{eqnarray}
\label{eq_eflx_r_stand}
 \left<F_r\right> = \frac{2\pi}{B^2}\left|\tilde{p}_{\rm tot, e}\right|^2 
    \mathrm{Re}\left(\frac{-i\omega}{\mu_{\rm e}^2}\frac{\tilde{p}'_{\rm tot, e}}{\tilde{p}_{\rm tot, e}}\right)
    \begin{cases}
      {\rm e}^{2\omega_{\rm I}t} & \quad \text{standing} \\
      {\rm e}^{-2 k_{\rm I} z} & \quad \text{propagating} .     
    \end{cases}
\end{eqnarray}
We note that $\left<F_r\right>$ for standing waves was originally derived in 
    \citeauthor{1986SoPh..103..277C}(\citeyear{1986SoPh..103..277C}, Eq.~(3.2)).
{\bf (See Appendix~A for a derivation of Eq.~(\ref{eq_eflx_r_stand}).)}    
Be the waves standing or propagating, $\tilde{p}_{\rm tot, e} \propto H_0^{(1)}(\mu_{\rm e}r)$ at large distances
    can be approximated by
\begin{eqnarray}
\label{eq_ptot_largeR}
   \tilde{p}_{\rm tot, e} \propto \sqrt{\frac{2}{\pi \mu_{\rm e}r}}\exp\left[i\left(\mu_{\rm e}r-\frac{\pi}{4}\right)\right] , 
\end{eqnarray}
    resulting in $\tilde{p}'_{\rm tot, e}/\tilde{p}_{\rm tot, e} \approx i\mu_{\rm e}$.
It then follows that
\begin{eqnarray}
\label{eq_eflx_r_stand_largeR}
   \left<F_r\right> \approx \frac{2\pi}{B^2}\left|\tilde{p}_{\rm tot, e}\right|^2
       \mathrm{Re}\left(\frac{\omega}{\mu_{\rm e}}\right)
    \begin{cases}
      {\rm e}^{2\omega_{\rm I}t} & \quad \text{standing} \\
      {\rm e}^{-2 k_{\rm I} z} & \quad \text{propagating} .     
    \end{cases}
\end{eqnarray}
From Fig.~\ref{fig_prop_vs_stand}b one can find that ${\rm Im}\left(\mu_{\rm e}\right) <0$ for leaky standing modes,
    meaning that $|\tilde{p}_{\rm tot, e}|$ tends to grow exponentially
    (Eq.~(\ref{eq_ptot_largeR}), barring the $r^{-1/2}$-dependence)
    and an eigen-mode analysis does not allow us to examine the energetics of the system.
However, this does not mean that this analysis is physically irrelevant because what matters is that
    the apparent temporal damping can be accounted for by the outwardly going energy flux density
    ($\left<F_r\right> >0$ because ${\rm Re}(\omega/\mu_e) >0$, see Eq.~(\ref{eq_eflx_r_stand_largeR})).
Indeed, numerical studies starting with initial standing waves suggest that the temporal damping
    after a transient stage matches exactly
    the attenuation rate given by the eigen-mode analysis~\citep{2007SoPh..246..231T}.
For propagating waves, one finds from Fig.~\ref{fig_prop_vs_stand}a that
    once again ${\rm Im}\left(\mu_{\rm e}\right) <0$ and ${\rm Re}(\omega/\mu_e) >0$.
Hence similar to the standing case, an eigen-mode analysis does not permit an investigation
    into the energetics of propagating waves.
However, if a coronal structure is perturbed with a harmonic boundary driver,
    one expects to see that the apparent spatial damping after some transient phase will be given
    by the attenuation length $1/k_{\rm I}$ obtained from this eigen-mode analysis.
And this attenuation is once again associated with the outwardly directed $\left<F_r\right>$.     
We note that this expectation can be readily numerically tested, in much the same way that the expected
    spatial damping of kink waves due to resonant absorption
    was tested~\citep{2013A&A...551A..40P}.

\section{Summary}
\label{sec_conc}

This study is motivated by the apparent lack of a dedicated study on the role that wave leakage plays in
    spatially attenuating propagating sausage waves supported by density-enhanced cylinders in the corona.
To this end, we worked in the framework of cold magnetohydrodynamics (MHD), 
   and numerically solved the dispersion relation~(DR, eq.~(\ref{eq_DR})) for complex-valued longitudinal 
   wavenumbers $k=k_{\rm R} + i k_{\rm I}$ at given real angular frequencies $\omega$.
To validate our numerical results, we also provided the analytical approximations to the full DR 
   in the low-frequency limit $\omega\rightarrow 0$
   and in the neighborhood of $\omega_{\rm c}$, the critical angular frequency separating
   trapped from leaky waves.
Our solutions indicate that while sausage waves can propagate for $\omega<\omega_{\rm c}$
   and can direct their energy upwards, they suffer substantial spatial attenuation.
The attenuation length ($1/k_{\rm I}$) is of the order of the cylinder radius $a$
   and shows little dependence on frequency or 
   the density contrast between a coronal structure and its surroundings
   for $\omega \lesssim 1.5 v_{\rm Ai}/a$, where $v_{\rm Ai}$ is
   the Alfv\'en speed in the cylinder.
This means that when a coronal cylinder is subject to boundary perturbations with a broadband spectrum
   (e.g., granular motions), wave leakage removes the low-frequency components rather efficiently.
A comparison with the solutions to the DR for standing waves (real $k$, complex $\omega$) indicates
   that a close relationship between propagating and standing waves
   exists only when the waves are trapped or weakly damped. 
In addition, while an eigen-mode analysis does not allow an investigation into the energetics of
   propagating waves, the attenuation length is expected to play an essential role 
   in numerical simulations where coronal structures are perturbed by harmonic boundary drivers.

%%%%%%%%%%%%%%%%%%%%%%%%%%%%%%%%%%%%%%%%%%%%%%%%%%%%%%%%%%%%%%%%%%%%%%%%%%
\begin{acknowledgements}
    This research is supported by the 973 program 2012CB825601, National Natural Science Foundation of China
    (41174154, 41274176, 41274178, and 41474149), 
    the Provincial Natural Science Foundation of Shandong via Grant JQ201212,
    and also by a special fund of Key Laboratory of Chinese Academy of Sciences.
\end{acknowledgements}

%%%%%%%%%%%%%%%%%%%%%%%%%%%%%%%%%%%%%%%%%%%%%%%%%%%%%%%%%%%%%%%%%%%%%%%%%%

\begin{appendix}
\section{A derivation of the averaged energy flux densities}
\renewcommand{\theequation}{\thesection\arabic{equation}}

This section offers a derivation of the averaged energy flux densities given in Eqs.~(\ref{eq_eflx_r}),
    (\ref{eq_eflx_z}) and (\ref{eq_eflx_r_stand}), following a procedure similar to that adopted by
    \citet[][page 1225]{2007ApJ...661.1222L}.
Let us first consider standing waves for which the axial wavenumber $k$ is real, but 
    the angular frequency $\omega$ is allowed to be complex-valued ($\omega = \omega_{\rm R} + i\omega_{\rm I}$).
Evaluating a perturbation $\delta f(r, z; t)$
    with Eq.~(\ref{eq_Fourier}) at given values of $[r, t]$ yields that
\begin{eqnarray}
\label{eq_f_breve_standing}
  \delta f(r, z; t) = {\rm Re}\left\{\breve{f}\exp(i kz)\right\} ,
\end{eqnarray}
    where 
\begin{eqnarray}
\breve{f} = \tilde{f}(r)\exp\left(-i\omega t\right) 
    = \tilde{f}(r)\exp\left(-i\omega_{\rm R} t\right){\rm e}^{\omega_{\rm I}t} .
\label{eq_brev_tilde}    
\end{eqnarray}
With another perturbation $\delta g(r, z; t)$ in the same form, one finds that
    the product $\delta f \delta g$ averaged over a wavelength $\lambda=2\pi/k$ is
\begin{eqnarray}
&& \left<\delta f\delta g\right> (r, t)		\nonumber \\
&\equiv & \frac{1}{\lambda}
    \int_0^{\lambda} \delta f(r, z;t) \delta g(r, z;t) {\rm d}z 	\nonumber \\
&=& \frac{1}{\lambda}
     \int_0^{\lambda} {\rm Re}\left\{\breve{f}\exp(i k z)\right\}
                      {\rm Re}\left\{\breve{g}\exp(i k z)\right\}
     {\rm d}z		\nonumber \\	 
&=& \frac{1}{\lambda}
    \int_0^{\lambda}\left[\breve{f}_{\rm R}\cos(kz)-\breve{f}_{\rm I}\sin(kz)\right] 
		    \left[\breve{g}_{\rm R}\cos(kz)-\breve{g}_{\rm I}\sin(kz)\right] 
    {\rm d}z		\nonumber \\	 
&=& \frac{1}{\lambda}
    \int_0^{\lambda}\left[\breve{f}_{\rm R}\breve{g}_{\rm R}\cos^2(kz)
			 +\breve{f}_{\rm I}\breve{g}_{\rm I}\sin^2(kz) \right.   \nonumber \\
& &		    \left. -\left(\breve{f}_{\rm R}\breve{g}_{\rm I}
			         +\breve{f}_{\rm I}\breve{g}_{\rm R}\right)\cos(kz)\sin(kz)
		    \right]
    {\rm d}z		\nonumber \\	 			 
&=& \frac{1}{2}\left(\breve{f}_{\rm R}\breve{g}_{\rm R}
		    +\breve{f}_{\rm I}\breve{g}_{\rm I}
	       \right)   \nonumber \\
&=& \frac{1}{2}{\rm Re}\left(\breve{f} \breve{g}^*\right)
    =\frac{1}{2}{\rm Re}\left(\breve{f}^* \breve{g}\right) ,
\label{eq_procedure_stand}    
\end{eqnarray}
    where we have used the shorthand notations $\breve{f}_{\rm R} = {\rm Re}\breve{f}$
    and $\breve{f}_{\rm I} = {\rm Im}\breve{f}$.
By noting that $\breve{f}$ is expressible in terms of $\tilde{f}$ through Eq.~(\ref{eq_brev_tilde}), 
    one finds that 
\begin{eqnarray}
\left<\delta f\delta g\right> (r, t)
   = \frac{1}{2}{\rm Re}\left[\tilde{f}(r)\tilde{g}^*(r)\right] {\rm e}^{2\omega_{\rm I} t}
   = \frac{1}{2}{\rm Re}\left[\tilde{f}^*(r)\tilde{g}(r)\right] {\rm e}^{2\omega_{\rm I} t} .
\end{eqnarray}
Now the energy flux density averaged over a wavelength can be evaluated with its definition, Eq.~(\ref{eq_eflx}),
    the results being
\begin{eqnarray}
&& \left< F_r \right> 
   = \frac{1}{2}\mathrm{Re}\left(\tilde{p}_{\rm tot} \tilde{v}_r^*\right) {\rm e}^{2 \omega_{\rm I} t}
   = \frac{1}{2}\mathrm{Re}\left(\tilde{p}_{\rm tot}^* \tilde{v}_r\right) {\rm e}^{2 \omega_{\rm I} t}, 
  \label{eq_eflx_r_prop} \\
&& \left< F_z \right> = \frac{1}{2}\mathrm{Re}\left( -\frac{B}{4\pi}\tilde{b}_r\tilde{v}_r^*\right) {\rm e}^{2 \omega_{\rm I} t} .
  \label{eq_eflx_z_prop}
\end{eqnarray}
Furthermore, by noting that Eq.~(\ref{eq_Fourier_amp}) allows $\tilde{v}_r$ to be expressed by 
\begin{eqnarray}
  \tilde{v}_r = -i\frac{\omega}{\mu^2} \frac{4\pi\tilde{p}_{\rm tot}'}{B^2} ,
\label{eq_vr_ptot_Famp}  
\end{eqnarray}
   one finds that
\begin{eqnarray}
  \left< F_r \right> 
&=& \frac{2\pi}{B^2}\mathrm{Re}\left[\tilde{p}_{\rm tot}^* \left(-i\frac{\omega}{\mu^2}\tilde{p}_{\rm tot}'\right)\right]
  {\rm e}^{2 \omega_{\rm I} t} ,  \nonumber \\
&=& \frac{2\pi}{B^2}\mathrm{Re}\left[\tilde{p}_{\rm tot}^* \tilde{p}_{\rm tot}
   \left(-i\frac{\omega}{\mu^2}\frac{\tilde{p}_{\rm tot}'}{\tilde{p}_{\rm tot}}\right)\right]
  {\rm e}^{2 \omega_{\rm I} t} ,  \nonumber \\
&=& \frac{2\pi \left|\tilde{p}_{\rm tot}\right|^2}{B^2}\mathrm{Re}
   \left(-i\frac{\omega}{\mu^2}\frac{\tilde{p}_{\rm tot}'}{\tilde{p}_{\rm tot}}\right)
       {\rm e}^{2 \omega_{\rm I} t} . 
\end{eqnarray}
This expression is valid both in and outside the cylinder.
When applied to the external medium, it results in the first expression in Eq.~(\ref{eq_eflx_r_stand}).

Now consider propagating waves for which $\omega$ is real, whereas $k$ is allowed to be complex-valued 
    ($k = k_{\rm R} + i k_{\rm I}$).
In this case evaluating a perturbation $\delta f(r, z; t)$
    with Eq.~(\ref{eq_Fourier}) at given values of $[r, z]$ yields that
\begin{eqnarray}
\label{eq_f_breve_prop}
  \delta f(r, z; t) = {\rm Re}\left\{\breve{f}\exp(-i\omega t)\right\} ,
\end{eqnarray}
    where 
\begin{eqnarray}
\breve{f} = \tilde{f}(r)\exp\left(i k z\right) 
    = \tilde{f}(r)\exp\left(i k_{\rm R} z\right){\rm e}^{-k_{\rm I}z} .
\label{eq_brev_tilde_prop}    
\end{eqnarray}
To evaluate the product $\delta f \delta g$ averaged over a wave period $T=2\pi/\omega$, 
    one can follow the same procedure as in Eq.~(\ref{eq_procedure_stand}) by replacing
    $kz$ with $-\omega t$, the result being
\begin{eqnarray}
&& \left<\delta f\delta g\right> (r, z) 		\nonumber \\
&\equiv& \frac{1}{T}
    \int_0^{T} \delta f(r, z; t)\delta g(r, z; t) {\rm d}t 	\nonumber \\
&=& \frac{1}{2}{\rm Re}\left(\breve{f} \breve{g}^*\right)
    =\frac{1}{2}{\rm Re}\left(\breve{f}^* \breve{g}\right) .
\end{eqnarray}
With the aid of Eq.~(\ref{eq_brev_tilde_prop}) which relates $\breve{f}$ to $\tilde{f}$,
    one finds that 
\begin{eqnarray}
\left<\delta f\delta g\right> (r, z)
 &=&  \frac{1}{2}{\rm Re}\left[\tilde{f}(r)\tilde{g}^*(r)\right] {\rm e}^{-2 k_{\rm I} z} \nonumber \\
 &=&  \frac{1}{2}{\rm Re}\left[\tilde{f}^*(r)\tilde{g}(r)\right] {\rm e}^{-2 k_{\rm I} z} .
\end{eqnarray}
The energy flux density averaged over a period follows from the definition, Eq.~(\ref{eq_eflx}),
    and the results are given by Eqs.~(\ref{eq_eflx_r}) and (\ref{eq_eflx_z}).
The second expression in Eq.~(\ref{eq_eflx_r_stand}), appropriate for propagating waves, 
    can be derived in view of Eq.~(\ref{eq_vr_ptot_Famp}). 
\end{appendix}
\end{document}